\documentstyle[12pt]{article}

\def\uglu{\hskip 0pt plus 1fil minus 1fil}
\def\uglux{\hskip 0pt plus .75fil minus .75fil}
\def\slashed#1{\setbox200=\hbox{$ #1 $} \hbox{\box200 \hskip
-\wd200 \hbox to \wd200 {\uglu $/$ \uglux}}}

\def\fH{{\cal H}}
\def\fL{{\cal L}}

\newcommand{\lsim}{\mathrel{\lower4pt\hbox{$\sim$}}
\hskip-12.5pt\raise1.6pt\hbox{$<$}\;}

\newcommand{\gsim}{\mathrel{\lower4pt\hbox{$\sim$}}
\hskip-12.5pt\raise1.6pt\hbox{$>$}\;}

\newcommand{\pl}{Phys.\ Lett.}
\newcommand{\np}{Nucl.\ Phys.}
\newcommand{\prl}{Phys.\ Rev.\ Lett.}

\newcommand{\etal}{{\it et al}.,\ }

\begin{document}

\setlength{\textheight}{7.5truein}

\def\vtd{$V_{td}$}
\def\what{\underbar{\hskip.5in}}

\def\lsim{\mathrel{\lower4pt\hbox{$\sim$}}
\hskip-12pt\raise1.6pt\hbox{$<$}\;}

\def\gsim{\mathrel{\lower4pt\hbox{$\sim$}}
\hskip-12pt\raise1.6pt\hbox{$>$}\;}

\begin{titlepage}
\noindent
\hspace*{10cm}SLAC-PUB-95-6962\\
\noindent
\hspace*{10cm}BNL- \\

\vfill
\begin{center}

{\bf Flavor Changing Neutral Scalar Currents
at $\mu^+\mu^-$ Colliders}\\

\vfill
{David Atwood$^a$,
Laura Reina$^b$
and  Amarjit Soni$^b$}\\
\end{center}
\vfill

\begin{flushleft}

a) Stanford Linear Accelerator Center,
Stanford University,
Stanford, CA\ \ 94309, USA \\
b) Department of Physics, Brookhaven National Laboratory,
Upton, NY\ \ 11973, USA\\

\end{flushleft}

\vfill

\begin{quote}

{\bf Abstract}: The prospect of observing the flavor changing decay
$\fH\to t\bar c$ of a neutral Higgs boson produced via $s$-channel
and its subsequent decay into $t\bar c$ is considered at a
$\mu^+\mu^-$ collider.  Numerical estimates are given
in the context of a two Higgs doublet model with flavor changing
couplings.  It is found that for many values of the model parameters
such
{\it tree-level} flavor changing
decays will be produced at an observable level. In addition studies
of the
helicity of the top will allow the determination of the relative
strengths of the flavor changing Higgs couplings and these may be measured
with about $10^3$ events.

\end{quote}

\vfill

\begin{center}

Submitted to {\it Physical Review Letters}
\end{center}

\vfill

\hrule
\vspace{5 pt}
\noindent
* This work was supported by US Department of Energy contracts
DE-AC03-765F00515 (SLAC)  and DE-AC02-76CH0016 (BNL).

\end{titlepage}

The suppression of flavor changing neutral currents (FCNC) is an
important feature of the Standard Model (SM).  Thus, the measurement of
such currents provides an important test which can discriminate
between the SM and various models of new physics. In the
SM, the relative largeness of the top mass \cite{cdf,d0}
leads to a measurable rate of FCNC's in the down type quark sector
through penguin processes \cite{bsga-the}. In fact recent experiments
at CLEO have observed the reaction $b\to s\gamma$. At least in part due to
the fact that no
correspondingly heavy down type quark is thought to exist, similar FCNC
processes within the up sector (e.g. $t\to c\gamma$) are
highly suppressed in the SM\cite{tcga-the}.
Since we do not know
of a
conservation law that enforces the absence of such FCNC's their
continual search is clearly warranted. These considerations have, of
course, fueled the searches for $\mu\to e\gamma$, $K_L\to\mu e$ etc.\
for a very long time. The extraordinary mass scale of the top quark has
prompted many to advocate that FCNC involving the top quark may
well
exist
\cite{hou}.

An important class of models where FCNC's can occur among up type
quarks are those where flavor changing occurs in an extended neutral
Higgs sector. In previous works \cite{luke,reina}, the observation of
FCNC's (due to penguin graphs involving such a Higgs sector) was
considered in the processes $t\to c \gamma$  or $cZ$ and $e^+ e^-$
(or indeed $\mu^+\mu^-$) $ \to \gamma$  or $Z\to t\bar c$
respectively.  In this Letter we suggest that the tree level coupling
of such flavor-changing neutral Higgs bosons \cite{lightcase} to $t\bar
c$ may be probed by
$\mu^+\mu^-\to t\bar c$ at suggested muon colliders (MUCs).

Although very much in the notion stage at present, the MUC has been
suggested
\cite{cline}-\cite{palmer}
as a possible lepton
collider for energies in the TeV range. The
advantage of such a MUC would be that the much heavier muon suffers
appreciably less energy loss from synchrotron and beamstrahlung
radiation. The obvious disadvantages include the fact that muons
eventually decay as well as the new accelerator technology development
needed to produce and control such beams to the necessary degree to
reach high luminosities.

If MUCs are eventually shown to be a practical and desirable tool for
exploring physics in the TeV range, most of the applications would be
very similar to electron colliders. One advantage however is that they
may be able to produce Higgs bosons directly in the $s$ channel in
sufficient quantity to study their properties directly
\cite{cline,barger,bargertwo,reina}. In particular, a simple but
fascinating possibility that we wish to explore here is when such a
Higgs, $\fH$,
has a flavor-changing  $\fH t\bar c$ coupling then the process
$\mu^+\mu^-\to t\bar c$ will give a signal which should be easy to
identify, is likely to take place at an observable rate and yet has a
negligible SM background. Thus the properties of this
important coupling can be studied in detail.

The crucial point is that in spite of the fact that the $\mu^+\mu^-\fH$
coupling, being
proportional to $m_\mu$, is very small, if the MUC is run on
the Higgs resonance, $\sqrt{s}=m_\fH$, Higgs bosons may be produced at
an appreciable rate \cite{cline,barger,bargertwo,reina}.

At $\sqrt{s}=m_\fH$, the cross section for producing $\fH$,
$\sigma_\fH$, normalized to $\sigma_0=\sigma(\mu^+\mu^-\to \gamma\to
e^+ e^-)$, is given by:

\begin{equation}
R(\fH) = {\sigma_\fH \over \sigma_0} = {3\over\alpha_e^2} B_\mu^\fH
\label{rdef}
\end{equation}

\noindent where $B_\mu^\fH$ is the branching ratio of $\fH\to \mu^+\mu^-$ and
$\alpha_e$ is the electromagnetic coupling.

If the Higgs is very narrow, the exact tuning to the resonance implied
in equation (\ref{rdef}) may not in general be possible. Let us suppose
then that the energy of the beam has a finite spread described by
$\delta$:

\begin{equation}
m_\fH^2(1-\delta)<s<m_\fH^2(1+\delta)
\end{equation}

\noindent where we assume that $s$ is uniform about this range. The effective
rate of Higgs production will thus be given by:

\begin{equation}
\tilde R(\fH)=\left[ \frac{\Gamma_\fH}{m_\fH\delta} \arctan
\frac{m_\fH\delta}{\Gamma_\fH} \right] R(\fH) \label{rtildef}
\end{equation}

We now consider an extended Higgs sector which admits FCNCs. In
refs.~\cite{luke,reina}, for instance, a minimal FCNC Higgs model with
two Higgs doublets $\phi_1,\phi_2$ is considered. We assume, without loss
of generality, that $\phi_1$ is aligned with the vev so that

\begin{equation}
<\phi_1>=\left(  \begin{array}{c}
0\\
{v/\sqrt{2}}
\end{array}
\right),\ \ \ \ <\phi_2>=0
\end{equation}

\noindent where $v=(\sqrt{2} G_F)^{-{1\over 2}}$.  There are three neutral mass
eigenstates  denoted by $H$, $h$, and $A$ which are \cite{luke,reina}

\begin{eqnarray}
H&=&\sqrt{2}[( {\rm Re} \phi_1^0 -v )\cos\alpha
+ {\rm Re} \phi_2^0 \sin \alpha ] \nonumber\\
h&=&\sqrt{2}[( -{\rm Re} \phi_1^0 -v )\sin\alpha
+ {\rm Re} \phi_2^0 \cos \alpha ] \nonumber\\
A&=&-\sqrt{2}{\rm Im} \phi^0_2
\end{eqnarray}

\noindent where the mixing angle $\alpha$ is a parameter determined by the
Higgs potential.

The Lagrangian for the Higgs-fermion interaction is \cite{luke,reina}:

\begin{eqnarray}
\fL &=& \lambda^U_{ij} \bar Q_i \tilde \phi_1 U_j + \lambda^D_{ij} \bar
Q_i \phi_1 D_j + \lambda^L_{ij} \bar L_i \phi_1 E_j \nonumber\\
&+& \xi^U_{ij}  \bar Q_i \tilde \phi_2 U_j  + \xi^D_{ij}  \bar Q_i \phi_2 D_j
+ \xi^L_{ij}  \bar L_i \phi_2 E_j \,\,\, + \mbox{h.c.}
\end{eqnarray}

\noindent Here the $\lambda^{U,D,L}_{ij}$ couplings turn out to be proportional
respectively to the quark and lepton mass matrices, while the
$\xi_{ij}$ couplings are arbitrary and flavor non-diagonal. For definiteness,
we will assume that the magnitude of the parameters $\xi_{ij}$ are as
suggested by the ansatz of \cite{sher},

\begin{equation}
|\xi_{ij}|\approx g {\sqrt{  m_im_j}\over M_w} \label{anza}
\end{equation}

Let us now consider that a Higgs $\fH$ of mass $m_\fH$ is under study at a
MUC\null.  For illustrative purposes we take $\fH=h$ in the above
model where $\alpha=0$ (case 1) or $\pi/4$ (case 2). The main
distinction between the two cases is that in case 2 the decays $\fH\to
ZZ,\ WW$ are possible while in case 1 they are not.  Thus case 1 is
very similar to $\fH=A$.  In general the coupling of $h$ to $f\bar f$
is:

\begin{equation}
C_{hff}=-{g\over 2}{m_f\over m_W}\sin\alpha+ { {\rm Re}\xi_{ff}+i\gamma_5
{\rm Im}\xi_{ff}\over \sqrt{2}} \cos\alpha \equiv {g m_f\over 2 m_W}
\chi_{f} e^{i\gamma_5\lambda_{f}}
\end{equation}

\noindent while the coupling to $ZZ$ and $WW$ is given by:

\begin{equation}
C_{hZZ}={g\sin\alpha\over\cos\theta_W }m_Z g^{\mu\nu} \ \ \ \
C_{hWW}={g\sin\alpha}m_W g^{\mu\nu}
\end{equation}

\noindent Finally the flavor changing Higgs $-t\bar c$ coupling is given by:

\begin{equation}
C_{htc}= {1\over\sqrt{2}}\left[ \xi_{tc}P_R+\xi_{ct}^\dagger P_L \right]
\cos\alpha \equiv {g\sqrt{m_t m_c}\over 2 m_W}(\chi_R P_R+\chi_L P_L)
\end{equation}

\noindent where $\chi_L$ and $\chi_R$ are in general complex numbers and of
order unity if (\ref{anza}) applies.

The decay rates to these modes given the above couplings can be
readily calculated at tree level by using the results that exist in
the literature \cite{gunion}:

\begin{eqnarray}
\Gamma(\fH\to t\bar t) &=& {3 g_W^2 m_t^2 m_\fH \over 32\pi m_W^2}
\beta_t\left[\beta_t^2+(1-\beta_t^2)\sin\lambda_t\right] \chi_t^2 \nonumber \\
\Gamma(\fH\to b\bar b) &=&  {3 g_W^2 m_b^2 m_\fH \over 32\pi m_W^2} \chi_b^2
\nonumber \\
\Gamma(\fH\to ZZ) &=& {g^2\over 128\pi}{m_\fH^3\over m_Z^2} \beta_Z
(\beta_Z^2+12{m_Z^4\over m_\fH^4})\sin^2\alpha \nonumber \\
\Gamma(\fH\to WW) &=& {g^2\over 64\pi}{m_\fH^3\over m_W^2} \beta_W
(\beta_W^2+12{m_W^4\over m_\fH^4})\sin^2\alpha
\end{eqnarray}

\noindent where $\beta_i=\sqrt{1-4m_i^2/m_\fH^2}$.

The decay rate to $t\bar c$ is thus:

\begin{equation}
\Gamma(\fH\to t\bar c)=  {3 g_W^2 m_t m_c m_\fH \over 32\pi m_W^2}
\left( {(m_\fH^2-m_t^2)^2\over m_\fH^4} \right) \left(
{|\chi_R|^2+|\chi_L|^2\over 2} \right)
\end{equation}

\noindent and, $\Gamma(\fH\to t\bar c)=\Gamma(\fH\to c\bar t)$ at the tree
level that we are considering for now. The decay rate to $\mu^+\mu^-$ which
we require in equation (\ref{rdef}) is

\begin{equation}
\Gamma(\fH\to \mu^+\mu^-) = \frac{g_W^2 m_\mu^2 m_\fH} {32\pi
m_W^2}\chi_\mu^2;   \ \ \ \  B^\fH_\mu=\Gamma(\fH\to \mu^+\mu^-)/\Gamma_\fH
\end{equation}

For the purpose of numerical estimates let us take
the following sample choices of parameters:

\begin{itemize}

\item{} Case 1:  $\alpha=\lambda_{c}=\lambda_{t}=0$, $\chi_\mu=\chi_b=\chi_t=1$
and $\chi_L=\chi_R=1$

\item{} Case 2:  $\alpha=\pi/4$, $\lambda_{c}=\lambda_{t}=0$,
$\chi_\mu=\chi_b=\chi_t=1$ and $\chi_L=\chi_R=1$

\end{itemize}

\noindent In figure~1 we plot $\tilde R_(\fH)$ with $\delta=0$, $10^{-3}$ and
$10^{-2}$ in the two cases as well as

\begin{equation}
\tilde R_{tc}= \tilde R(\fH)\,(B^\fH_{t\bar c}+B^\fH_{c\bar t})
\end{equation}

\noindent Note that in case~1 if $m_\fH$ is below the $t\bar t$ threshold
$\tilde R_{tc}$ is about $.01-1$ and in fact $tc$ makes up a large
branching ratio.  Above the $t\bar t$ threshold $R_{tc}$ drops. For
case 2 the branching ratio is smaller due to the $WW$ and $ZZ$
threshold at about the same mass as the $tc$ threshold and so $R_{tc}$
is around $10^{-3}$. For a specific example if $m_\fH=300 GeV$,
then $\sigma_0\approx 1pb$. For a  luminosity of $10^{34}
cm^{-2} s^{-1}$, a year of $10^7 s$ ($1/3$ efficiency) and
for $\delta=10^{-2}$ case 1 will produce about $5\times10^3 (t\bar
c+\bar t c)$
events and case~2 will produce about $150$ events. Given the
distinctive nature of the final state and the lack of a Standard Model
background, sufficient luminosity should allow the observation of such
events.

If such events are observed one would like to extract the values of
$\chi_L$ and $\chi_R$. What is measured initially at a $\mu^+\mu^-$
collider is $\tilde R_{tc}$. One is required to know the total width
of the $\fH$ and the energy spread of the beam in order to translate
this into $\Gamma(\fH\to  t\bar c)$. This then allows the
determination of $|\chi_L|^2+|\chi_R|^2$. To get information
separately on the two couplings we note that the total helicity of the
top is:

\begin{equation}
{\bf H}_t=-{\bf H}_{\bar t}= {|\chi_R|^2-|\chi_L|^2  \over
|\chi_R|^2+|\chi_L|^2}
\end{equation}

\noindent from which one may therefore infer $|\chi_L|$ and
$|\chi_R|$. Unfortunately in the limit of small $m_c$ the helicity of
the $c$-quark is conserved hence the relative phase of $\chi_L$ and
$\chi_R$ may not be determined since the two couplings do not
interfere.

Of course the helicity of the $t$ cannot be observed directly, however
following the discussion of \cite{mumu} one may obtain it from the
decay distributions of the top. In particular if $X$ is a particle
arising in top decay let us define the forward-backwards asymmetry

\begin{equation}
A_X= { \Gamma(\cos\theta_X>0)-\Gamma(\cos\theta_X<0)
\over\Gamma(\cos\theta_X>0)+\Gamma(\cos\theta_X<0) }
\end{equation}

\noindent where $\theta_X$ is the angle between $\vec P_X$ and
$-\vec P_\fH$ in the $t$ rest frame.
For each particular choice of $X$ we define $\epsilon_X$ to be the
correlation with the polarization defined by:

\begin{equation}
\epsilon_X=3<\cos\theta^t_X>
\end{equation}

\noindent where $\theta^t_X$ is the angle between $X$ and the spin axis of a
polarized top.

In terms of $\epsilon_X$ the asymmetry $A_X$ is thus given by:

\begin{equation}
A_X= {1\over 2} \epsilon_X  {\bf H}_t.
\end{equation}

\noindent Let us now consider the following  decays \cite{mumu}:

\begin{itemize}

\item{} 1) for $t\to Wb, W\to l^+\nu_l$ where $l=e$, $\mu$
then  $\epsilon_l=1$
and the branching fraction for this
case is $B_1\sim\frac{2}{9}$.

\item{} 2) For $t\to Wb, W\to{}$hadrons then
$\epsilon_W= (m_t^2-2m_W^2)/(m_t^2+2m_W^2) \approx 0.39$
and the branching fraction for this is $B_2\sim\frac{7}{9}$.

\end{itemize}

The number of $t\bar c$ events needed to observe
the top helicity with a significance of 3-$\sigma$ is \cite{mumu}:

\begin{equation}
N_{3\sigma}={36\over{\cal E}_t^2{\bf H}_t^2} \approx {107\over {\bf H}_t^2}
\end{equation}

\noindent where

\begin{equation}
{\cal E}_t=\sqrt{B_1\epsilon_l^2+B_2\epsilon_W^2}\approx.58
\end{equation}

\noindent Thus at least $10^2$ events are required to begin to measure the
helicity of the top and hence the relative strengths of $\chi_L$ and
$\chi_R$.  In the above numerical examples it is clear that for some
combinations of parameters, particularly if the luminosity is
$10^{34}cm^{-2}s^{-1}$, sufficient events to measure the helicity may be
present.

\bigskip
\bigskip
\bigskip

This work was supported by US Department of Energy contracts
DE-AC03-765F00515 (SLAC)  and DE-AC02-76CH0016 (BNL).

\pagebreak

\bigskip
\noindent{\bf Figure Captions}
\medskip

\bigskip
\medskip
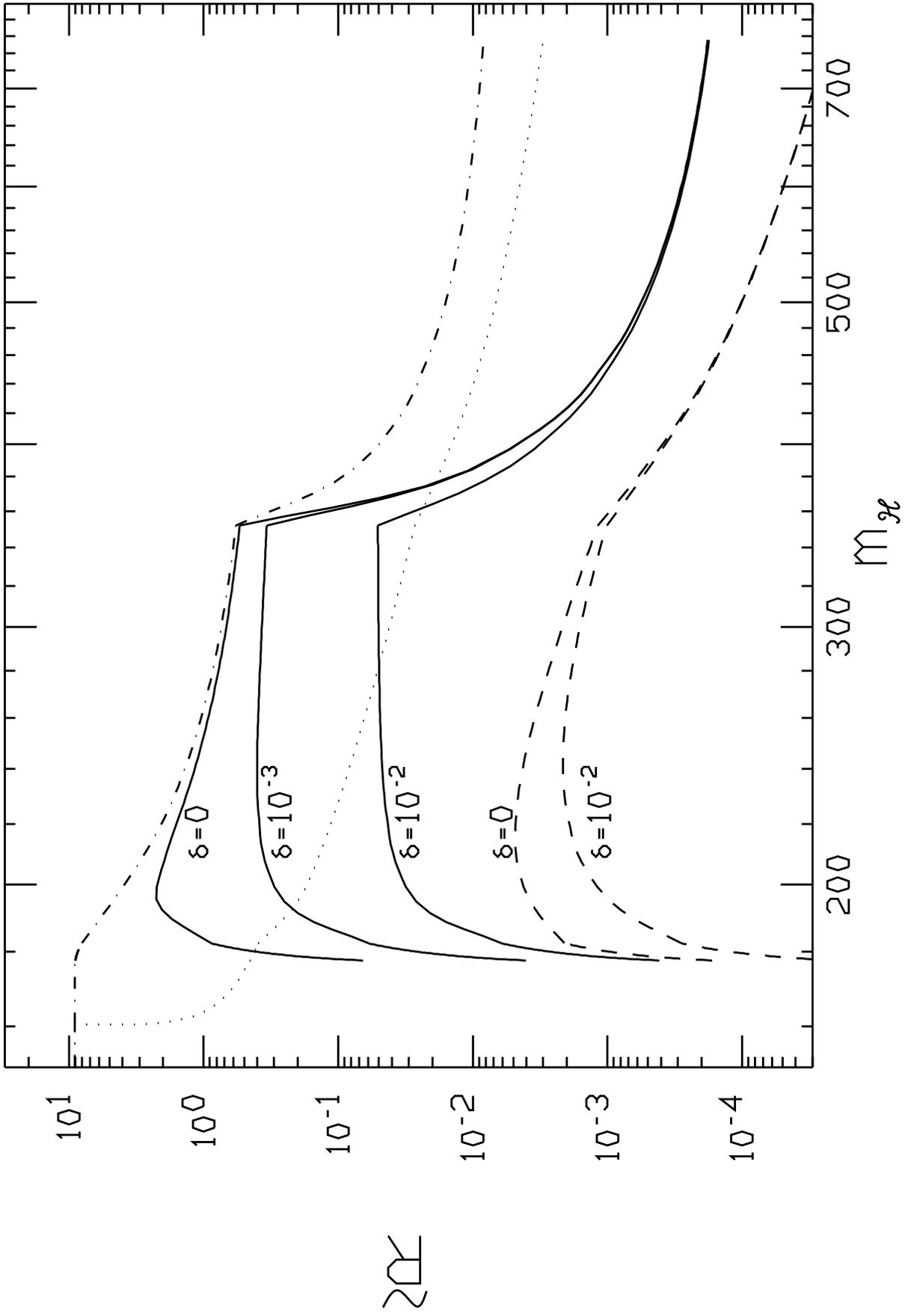
\noindent{\bf Figure 1}:
The value of $\tilde R(\fH)$ is shown  as a function of $m_{\fH}$
for scenario 1 (dash-dot) and  for scenario 2 (dots). The value of $\tilde
R_{tc}$ is shown in case 1 for
$\delta=0$ (upper solid curve); $\delta=10^{-3}$ (middle solid curve)
and $\delta=10^{-2}$ (lower  solid curve). The value of $\tilde R_{tc}$ is
shown
in case 2 for $\delta=0$ (upper dashed curve)
and $\delta=10^{-2}$ (lower dashed curve).

\pagebreak


\begin{thebibliography}{99}

\bibitem{cdf} F. Abe \etal [CDF], \prl\ {\bf74}, 2626 (1995);

\bibitem{d0} S. Abachi \etal [$D\emptyset$], \prl\ {\bf74},2632 (1995).

\bibitem{bsga-the} B. Grinstein, R. Springer, and M. Wise, Phys.\
Lett.\ {\bf B202}, 132 (1988); Nucl.\ Phys.\ {\bf B339}, 269 (1990);
R. Grigjanis, P.J. O'Donnel, M. Sutherland and H. Navelet,
\pl\ {\bf B213}, 355 (1988); \pl\ {\bf B223}, 239 (1989);
\pl\ {\bf B237}, 252 (1990);
G. Cella, G. Curci, G. Ricciardi and A. Vicer\'e, \pl\ {\bf B248},
181 (1990), \pl {\bf B325}, 227 (1994);
M.~Misiak, Nuc. Phys {\bf B393}, 23 (1993);
A.J. Buras, M. Misiak, M. Munz,
and S. Pokorski, Nuc. Phys. {\bf B424}, 374 (1994);
K. Adel and Y.P. Yao,
Phys. Rev. {\bf D49}, 4945 (1994);
M. Ciuchini, E. Franco, M.Martinelli, L. Reina, L. Silvestrini,
\pl\ {\bf B 316}, 127 (1993); \np\ {\bf B421}, 4 (1994);
\pl\ {\bf B 344}, 137 (1994).

\bibitem{tcga-the}
V. Ganapathi, T. Weiler, E. Laermann, I. Schmitt, and P.M. Zerwas,
Phys.\ Rev.\ {\bf D27}, 579 (1983); A. Axelrod, Nucl.\ Phys.\ B{\bf209}
349 (1982); G. Eilam, J.L. Hewett and A. Soni, Phys.\ Rev.\ {\bf D44},
1473 (1991); B. Grzadkowski, J.F. Gunion, and P. Krawczyk, Phys.\
Lett.\ {\bf B268}, 106 (1991).

\bibitem{hou}
M.J. Savage, Phys.\ Lett.\ {\bf B266},135 (1991);
W.S. Hou, Phys.\ Lett.\ {\bf B296}, 179 (1992);
M. Luke and M. J. Savage, Phys.\ Lett.\  {\bf B307}, 387 (1993);
L.J. Hall and S.Weinberg, Phys.\ Rev.\ {\bf D48}, R979 (1993);
G.C. Branco, P.A. Parada, M.N. Rebelo, UWTHPH-1994-51, (1995);
T. Han, R.D. Peccei, and X. Zhang, preprint FERMILAB-PUB-95/160-T;\ \ \
D. Atwood, L. Reina, and A. Soni, hep-ph/9506243.

\bibitem{luke}  M. Luke and M. J. Savage, Ref.~5.

\bibitem{reina}
D.~Atwood, \etal Ref.~5.

\bibitem{lightcase} In the case the Higgs is lighter than the $t$, one
has $t\to cH^0$ as emphasized by W.S. Hou,
see Ref.~5.

\bibitem{cline} D. Cline, Nucl.\ Instr.\ \& Meth.\ A{\bf350}, 24 (1994).

\bibitem{neuffer} D.V. Neuffer, {\it ibid}, p.~27.

\bibitem{barletta} W.A. Barletta and A.M. Sessler, {\it ibid}, p.~36;
A.G. Ruggiero, {\it ibid}, p.~45; S. Chattopadhyay {\it et al}., {\it
ibid}, p.~53.

\bibitem{palmer} R.B. Palmer, preprint, ``High Frequency $\mu^+ \mu^-$
Collider Design,'' SLAC-AAS-Note81, 1993.

\bibitem{barger} V. Barger, M. Berger, K. Fujii, J.F. Gunion, T. Han, C.
Heusch, W. Hong, S.K. Oh, Z. Parsa, S. Rajpoot, R. Thun and B. Willis,
Working group report, {\it Proc.\ of the First Workshop on the Physics
Potential and Development of $\mu^+\mu^-$ Colliders, Sausalito, CA},
preprint MAD-PH-95-873 (BB-9503258).

\bibitem{bargertwo} V. Barger, M. Berger, J. Gunion, and T. Han,
hep-ph/9404330.

\bibitem{sher} T.P. Cheng and M. Sher, Phys.\ Rev.\ {\bf D35},
3484 (1987);
M. Sher and Y. Yuan, Phys.\ Rev.\ {\bf D44}, 1461 (1991).

\bibitem{gunion} J. Gunion, H. Haber, G. Kane and S. Dawson, ``The Higgs
Hunters Guide,''  published by Addison Wesley.

\bibitem{mumu}
D.~Atwood and A.~Soni, SLAC-PUB-95-6877.

\end{thebibliography}
\end{document}